\documentclass[11pt,a4paper]{article}

\usepackage[utf8]{inputenc}
\usepackage[T1]{fontenc}
\usepackage{authblk}
\usepackage{epsfig}
\usepackage{subcaption}
\usepackage{calc}
\usepackage{amssymb}
\usepackage{amstext}
\usepackage{amsmath}
\usepackage{amsthm}
\usepackage{multicol}
\usepackage{algorithm2e}
\usepackage[bottom]{footmisc}
\usepackage{booktabs}
\usepackage{array}
\usepackage{graphicx}
\usepackage{color}
\usepackage{tabularx}
\usepackage{breqn}
\usepackage{enumitem}
\usepackage{natbib}
\usepackage[colorlinks=true,linkcolor=blue,citecolor=blue,urlcolor=blue]{hyperref}
\graphicspath{{./img/}{./}}

\newcommand{\Ntotal}{2{,}008{,}246}

\title{A CEFR-Inspired Classification Framework with Fuzzy C-Means To Automate Assessment of Programming Skills in Scratch}

\author[1]{Ricardo Hidalgo-Aragón}
\author[1]{Jesús M. González-Barahona}
\author[1]{Gregorio Robles}

\affil[1]{Universidad Rey Juan Carlos, Madrid, Spain}
\affil[ ]{\small \texttt{\{r.hidalgoa.2024@alumnos, jesus.gonzalez.barahona, grex@gsyc\}@urjc.es}}

\date{}

\begin{document}

\maketitle

\begin{abstract}
\noindent\textit{Context:} Schools, training platforms, and technology firms increasingly need to assess programming proficiency at scale with transparent, reproducible methods that support personalized learning pathways. \textit{Objective:} This study introduces a pedagogical framework for Scratch project assessment, aligned with the Common European Framework of Reference (CEFR), providing universal competency levels for students and teachers alongside actionable insights for curriculum design. \textit{Method:} We apply Fuzzy C-Means clustering to \Ntotal{} Scratch projects evaluated via Dr.Scratch, implementing an ordinal criterion to map clusters to CEFR levels (A1--C2), and introducing enhanced classification metrics that identify transitional learners, enable continuous progress tracking, and quantify classification certainty to balance automated feedback with instructor review.
\textit{Impact:} The framework enables diagnosis of systemic curriculum gaps---notably a ``B2 bottleneck'' where only 13.3\% of learners reside due to the cognitive load of integrating Logic, Synchronization, and Data Representation---while providing certainty---based triggers for human intervention.
\end{abstract}


\section{Introduction}
\label{sec:intro}

Computational Thinking~\citep{lodi_computational_2021} (CT) is a key skill for the digital age, involving the ability to solve complex problems using core computing concepts like decomposition, abstraction, logic, and flow control. Recognizing its importance, many educational programs worldwide have incorporated CT into school curricula, teacher training, and informal learning. The connection between programming abilities and CT is essential, carrying important consequences for educational and career advancement~\citep{ROMANGONZALEZ2017678}.

Beyond measuring CT, educators need frameworks that provide \textit{actionable guidance}---tools that not only classify learners but also reveal where students struggle, which skills to prioritize at each stage, and when automated feedback suffices versus when instructor intervention is warranted.

Visual, block-based programming tools, especially Scratch\footnote{\url{https://scratch.mit.edu/}}, are commonly used to teach CT, introducing learners to programming basics through creative, hands-on problem-solving. Several tools now exist that can help quantify CT-related skills, such as Dr.Scratch\footnote{\url{https://www.drscratch.org/}}~\citep{moreno2015dr,moreno2014automatic}.

Several automated tools have emerged to assess computational thinking in block-based environments. Beyond Dr.Scratch, Hairball~\citep{moreno2014automatic} detects code smells and anti-patterns in Scratch projects, while LitterBox~\citep{LitterBox_linter} identifies bugs and quality issues through static analysis. These tools produce valuable diagnostic information but rely on discrete metrics---counts of specific constructs or binary presence/absence indicators---without providing an integrated proficiency classification. Complementary assessment approaches include the Computational Thinking Test (CTt)~\citep{romangonzalez2015ctt,ROMANGONZALEZ2017678}, which measures CT through multiple-choice items validated against cognitive ability tests, and competition-based assessments like Bebras\footnote{\url{https://www.bebras.org/}} that evaluate algorithmic reasoning independently of programming artifacts. A systematic mapping~\citep{alves2019approaches} identified over 40 approaches to CT assessment in K-12 education, revealing substantial methodological diversity but limited standardization across proficiency scales.
The Common European Framework of Reference for Languages (CEFR)~\citep{council2001common} provides a widely recognized six-level proficiency scale (A1--C2) that has been adopted beyond natural languages. The European Framework for the Digital Competence of Educators (DigCompEdu)~\citep{ghomi2019digital,punie2017european} employs an analogous six-level structure to certify teachers' digital skills across 22 competencies. More directly relevant to programming is pycefr~\citep{robles_pycefr_2022}, a tool that classifies Python code snippets according to the language constructs required for comprehension, mapping them to CEFR-inspired levels. This precedent demonstrates the viability of adapting proficiency frameworks originally designed for natural languages to programming contexts. We extend this approach by combining CEFR-aligned classification with fuzzy clustering, enabling not only discrete level assignment but also quantification of transitional states between adjacent proficiency bands.
Fuzzy C-Means (FCM)~\citep{bezdek1984fcm} clustering offers unique advantages for educational assessment compared to hard clustering alternatives. Unlike K-Means~\citep{macqueen1967}, which forces each observation into exactly one cluster, FCM assigns membership degrees across all clusters, naturally representing learners who exhibit characteristics of multiple proficiency levels simultaneously. This soft partitioning aligns with established models of skill acquisition~\citep{dreyfus1986mind} that characterize learning as gradual progression through overlapping stages rather than discrete jumps. FCM has demonstrated effectiveness in educational contexts, including learner profiling in e-learning systems~\citep{hogo_evaluation_2010} and student performance prediction. Density-based approaches like DBSCAN~\citep{ester1996dbscan}, while effective for discovering arbitrary cluster shapes, produce fragmented outputs when applied to broadly distributed educational data and cannot impose the ordinal structure required for proficiency-level classification.

However, while these existing tools and frameworks address specific aspects of CT assessment, the resulting models often lack a coherent framework that classifies learners into several increasingly complex performance levels. Such a framework would enable classifying projects into clusters of complexity, offering a learning path analogous to the one used to accredit the Digital Competence of Educators~\citep{ghomi2019digital} in the European framework.

To achieve this, this paper presents a new method for categorizing Scratch projects by complexity using FCM clustering~\citep{ross_timothy_j_fuzzy_2010}, whose soft partitioning naturally captures the gradual, overlapping nature of skill acquisition---a critical consideration for educational settings where forcing students into rigid categories obscures their developmental state.


\section{Methodology}
\label{sec:meth}

\subsection{Data Collection and Preparation}

\subsubsection{Dataset Origin and Scope}

The empirical foundation comprises \textbf{\Ntotal{} public Scratch projects} retrieved from the official Scratch repository through software mining techniques. Each project was automatically evaluated using Dr.Scratch~\citep{moreno2015dr}, an established automated assessment tool that quantifies CT proficiency across nine dimensions. This dataset represents one of the largest empirical studies of programming skill assessment to date.

All data originates from publicly shared Scratch projects published under a Creative Commons CC BY-SA 2.0 license. To ensure privacy protection, only the nine aggregated CT metrics generated by Dr.Scratch were retained; no source code, media assets, or personally identifiable information were stored. The complete dataset on which this study is based is available via Kaggle~\citep{kaggle_drscratch_metrics}.

\subsubsection{Computational Thinking Dimensions}

Each Scratch project is characterized by a nine-dimensional feature vector $\mathbf{x}_i \in \{0,1,2,3,4\}^9$, where each component represents an ordinal assessment (ranging from 0 to 4) of the following CT dimensions as defined by Dr.Scratch: Abstraction (use of custom blocks, cloning, procedural decomposition), Parallelization (concurrent execution, event-driven programming), Logic (Boolean operators, conditionals), Synchronization (coordination between sprites/scripts), Flow Control (loops, iteration structures), User Interactivity (event handling, user input), Data Representation (variables, lists, data structures), Math Operators (arithmetic operations), and Motion Operators (sprite positioning, movement commands).

These dimensions collectively capture both the breadth and depth of computational thinking skills, providing a multidimensional profile suitable for fuzzy classification.

\subsubsection{Data Characteristics and Train-Test Split}

Descriptive statistics (Table~\ref{tab:descriptive_stats}) reveal substantial heterogeneity in skill distributions. Several dimensions exhibit pronounced class imbalance: Math Operators and Logic show heavy concentration in lower categories (0--1), reflecting their advanced nature, while Flow Control and Data Representation demonstrate more uniform distributions. This naturally occurring imbalance mirrors authentic learning contexts.

\begin{table}[ht]
	\centering
	\caption{Descriptive statistics for the complete dataset (N = \Ntotal{})}
	\label{tab:descriptive_stats}
	\small
	\begin{tabular}{lccccc}
		\toprule
		\textbf{Dimension} & \textbf{Mean} & \textbf{SD} & \textbf{Q1} & \textbf{Med} & \textbf{Q3} \\
		\midrule
		Abstraction & 1.47 & 1.36 & 1 & 1 & 3 \\
		Parallelization & 1.75 & 1.60 & 0 & 1 & 4 \\
		Logic & 1.26 & 1.60 & 0 & 0 & 3 \\
		Synchronization & 1.54 & 1.42 & 0 & 1 & 3 \\
		Flow Control & 2.07 & 1.22 & 1 & 2 & 3 \\
		User Interactivity & 1.55 & 0.82 & 1 & 2 & 2 \\
		Data Representation & 2.10 & 1.60 & 1 & 1 & 4 \\
		Math Operators & 0.76 & 1.33 & 0 & 0 & 1 \\
		Motion Operators & 1.80 & 1.62 & 0 & 2 & 3 \\
		\bottomrule
	\end{tabular}
\end{table}

The complete dataset was partitioned into training (80\%, $n=1{,}606{,}596$) and test sets (20\%, $n=401{,}650$) using stratified random sampling. The training set was used for: (1) FCM clustering and centroid determination, (2) cluster ordering via $S_j$ criterion, (3) statistical validation, and (4) 5-fold cross-validation. The held-out test set evaluated generalization performance, enhanced classification efficacy, and comparison against baseline algorithms.

\subsection{Fuzzy C-Means Clustering for Ordinal Classification}

Traditional hard clustering algorithms (e.g., K-Means~\citep{macqueen1967} or DBSCAN~\citep{ester1996dbscan}) force each observation into a single discrete category, thereby discarding valuable information about degrees of membership to multiple skill levels. In educational contexts, where learners often exhibit transitional competencies spanning adjacent proficiency bands, this binary assignment proves inadequate.

Fuzzy C-Means (FCM)~\citep{bezdek1984fcm} addresses this limitation by assigning each project a membership vector $\mathbf{u}_i = [u_{i1}, u_{i2}, \ldots, u_{ic}]$ where $u_{ij} \in [0,1]$ quantifies the degree to which project $i$ belongs to cluster $j$, subject to the constraint $\sum_{j=1}^{c} u_{ij} = 1$. This soft partitioning naturally accommodates the gradual, overlapping nature of skill acquisition.

To determine optimal algorithm parameters, a grid search was conducted over the parameter space $m \in \{1.5, 2.0, 2.5\}$ and $\epsilon \in \{10^{-3}, 10^{-4}, 10^{-5}\}$. The configuration maximizing both Fuzzy Partition Coefficient (FPC) and Silhouette Score was selected: fuzzification parameter $m = 1.5$ (yielding crisper cluster boundaries appropriate for ordinal skill levels) and convergence tolerance $\epsilon = 10^{-5}$ (ensuring high-precision centroid convergence), with maximum iterations set to 1000.

\subsection{Ordinal Cluster Ordering via $S_j$ Criterion}
\label{sec:sj_ordering}

FCM inherently produces unordered clusters---the algorithm optimizes spatial compactness without imposing any ordinality constraint. However, when applied to ordinal variables (CT scores ranging 0--4), a natural ordering emerges through the aggregate membership distribution. We introduce the \textbf{$S_j$ criterion}, which leverages the ordinal nature of input features to derive a data-driven ranking of clusters.

For each cluster $j$, we define the total centroid sum:
\begin{equation}
	S_j = \sum_{d=1}^{9} v_{jd}
	\label{eq:sj_criterion}
\end{equation}
where $v_{jd}$ is the value of centroid $j$ in dimension $d$. Since higher CT scores indicate greater proficiency, clusters with larger $S_j$ values correspond to more advanced skill profiles. By sorting clusters in ascending order of $S_j$, we obtain an ordinal mapping to CEFR levels (A1, A2, B1, B2, C1, C2).

\subsubsection{Justification for Six Clusters}

\begin{figure}[htbp]
	\centering
	\includegraphics[width=0.75\linewidth]{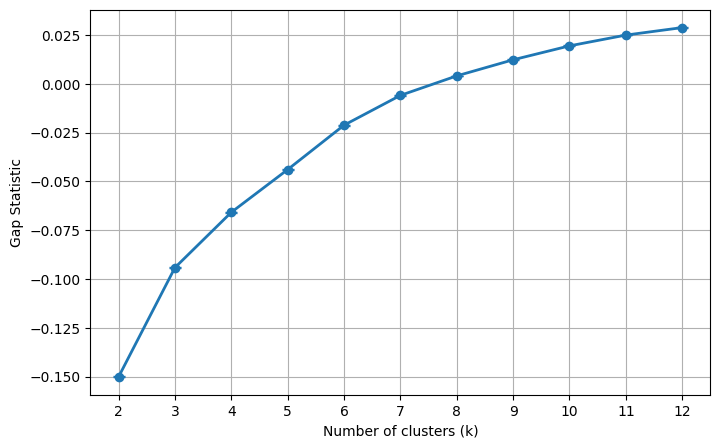}
	\caption{Gap statistic for $k = 2$ to $k = 12$ clusters. Monotonic increase without a clear peak and negative values for $k \leq 6$ reflect the continuous nature of programming skill distributions.}
	\label{fig:gap_statistic}
\end{figure}

Determining the optimal number of clusters presents a tension between statistical parsimony and pedagogical utility. Internal validity indices consistently favor fewer clusters: the Gath--Geva index~\citep{gath2002unsupervised} is optimal at $k=2$, Fuzzy Silhouette Width~\citep{rousseeuw1987silhouettes} drops from 0.256 ($k=2$) to 0.013 ($k=6$), and the Fuzzy Dunn Index~\citep{dunn1973fuzzy} decreases from 0.64 to 0.41. The Gap statistic~\citep{tibshirani2001gap,yue_extension_2013} exhibits monotonic increase across $k \in [2, 12]$ without a clear elbow (Figure~\ref{fig:gap_statistic}), with negative values for $k \leq 6$ indicating that cluster structure is less compact than uniformly distributed data---a pattern characteristic of broadly overlapping skill distributions.

Despite these statistical preferences, we adopt $k=6$ based on three complementary justifications:

\textbf{Domain alignment.} The six-level CEFR structure is internationally recognized and already adopted for educator digital competence via DigCompEdu~\citep{ghomi2019digital}, facilitating integration into existing certification infrastructure. A binary partition ($k=2$) would collapse meaningful variance, rendering the model blind to the 13.7\% of transitional learners---the very population where pedagogical intervention is most critical.

\textbf{Statistical discrimination.} Although clusters overlap substantially, all pairwise comparisons remain statistically significant (Kruskal--Wallis $H = 7{,}609{,}964.74$, all Mann--Whitney U tests $p < 0.001$). Dimension-specific rank correlations confirm monotonic ordering across levels (Synchronization $\tau = 0.966$, Flow Control $\tau = 0.894$).

\textbf{Enhanced classification utility.} The fuzzy membership framework captures 13.7\% of observations as transitional states between adjacent levels---information that would be lost with hard clustering or coarser granularity.

This principled trade-off---sacrificing maximal cluster compactness for actionable granularity---aligns with the recognition that skill progression manifests as overlapping distributions rather than discrete categories.

\subsection{Empirical Threshold Selection}

The classification thresholds were selected through systematic sensitivity analysis examining how threshold variations affect classification distributions and pedagogical utility.

The combined threshold configuration ($\tau_{clear}=0.5$, $\tau_{trans}=0.15$) yields a pedagogically interpretable distribution: 79.1\% Clear, 13.7\% Transition, and 7.3\% Predominant classifications. Heatmap analysis of the joint parameter space confirms that this configuration lies within a stable region where small parameter perturbations produce proportionate (not abrupt) changes in classification outcomes.

\subsection{Enhanced Classification System}

Traditional \texttt{argmax} classification---which assigns each observation to the single cluster with highest membership---discards valuable information encoded in the full membership vector. More critically for educators, it forces transitional learners---those actively developing skills between adjacent levels---into discrete categories that misrepresent their actual learning state. To address this pedagogical limitation, we introduce three complementary metrics with empirically grounded thresholds, each designed to support specific educational decisions.

\paragraph{Classification Type}

We categorize observations into three types based on membership distribution characteristics:

\begin{equation}
	\small
	\text{Type}(\mathbf{u}_i) =
	\begin{cases}
		\text{Clear} & \text{if } \max(\mathbf{u}_i) \geq 0.5 \\
		\text{Transition} & \text{if } |u_{i,\text{primary}} - u_{i,\text{secondary}}| < 0.15 \\
		\text{Predominant} & \text{otherwise}
	\end{cases}
	\label{eq:classification_type}
\end{equation}

The threshold $\tau_{\text{clear}} = 0.5$ ensures that \textit{Clear} assignments reflect genuine majority membership---the dominant cluster accounts for more than all others combined. The transition threshold $\tau_{\text{trans}} = 0.15$ captures cases where primary and secondary memberships are sufficiently close to indicate active skill development between adjacent levels. This value corresponds approximately to one standard deviation of the membership difference distribution in our dataset, isolating the upper tail of ambiguous classifications. Observations falling between these criteria are labeled \textit{Predominant}: a discernible primary cluster exists, but without absolute majority.

\paragraph{Continuous Score}

Rather than discrete level assignment, we compute a continuous proficiency indicator $S_{\text{cont}} \in [1, 6]$ as the membership-weighted average of CEFR ordinals:

\begin{equation}
	S_{\text{cont}}(\mathbf{u}_i) = \sum_{j=1}^{6} u_{ij} \cdot \ell_j
	\label{eq:continuous_score}
\end{equation}

\noindent where $\ell_j \in \{1, 2, 3, 4, 5, 6\}$ maps to levels \{A1, A2, B1, B2, C1, C2\}. This metric enables fine-grained progress tracking: a learner advancing from $S_{\text{cont}} = 2.3$ to $S_{\text{cont}} = 2.7$ demonstrates measurable growth even without crossing a discrete level boundary.

\paragraph{Certainty Quantification}

Classification confidence is measured using normalized Shannon entropy~\citep{shannon1948mathematical}:

\begin{equation}
	\text{Certainty}(\mathbf{u}_i) = 1 - \frac{H(\mathbf{u}_i)}{H_{\max}} = 1 - \frac{-\sum_{j=1}^{6} u_{ij} \log_2(u_{ij})}{\log_2(6)}
	\label{eq:certainty}
\end{equation}

\noindent where $H_{\max} = \log_2(6) \approx 2.585$ represents maximum entropy (uniform distribution across all six clusters). Certainty ranges from 0 (completely ambiguous) to 1 (deterministic).

We partition certainty into three interpretable levels using thresholds derived from the entropy scale. \textit{Low} certainty ($< 0.4$) corresponds to entropy exceeding 60\% of maximum, indicating substantial ambiguity where instructor review may be warranted. \textit{High} certainty ($\geq 0.7$) corresponds to entropy below 30\% of maximum, reflecting confident classifications suitable for automated feedback. The intermediate \textit{Medium} range captures typical cases requiring contextual interpretation.

This certainty quantification serves a critical pedagogical function: it operationalizes the balance between automated and human-mediated assessment. Classifications with \textit{Low} certainty signal cases where the algorithm's confidence is insufficient for autonomous feedback---these learners may benefit from instructor review, portfolio analysis, or supplementary assessment. Conversely, \textit{High} certainty classifications are suitable for automated formative feedback, enabling scalable assessment without sacrificing accuracy. This design embodies a ``human-in-the-loop'' approach increasingly recognized as best practice in educational technology~\citep{hahn2021systematic}.

\subsection{Validation Strategy}

The generalization of the model is assessed by comparing Silhouette Score~\citep{rousseeuw1987silhouettes}, Fuzzy Partition Coefficient (FPC), Partition Entropy, and Average Certainty across training versus held-out test sets. Five-fold cross-validation on the training set confirms stability across random data partitions.

Three complementary statistical tests verify ordinal progression: (1) Kruskal--Wallis H-test~\citep{kruskal1952ranks} for global distribution differences; (2) Pairwise Mann--Whitney U tests~\citep{mann1947test} for adjacent CEFR levels; (3) Kendall's $\tau$~\citep{kendall1938new} and Spearman's $\rho$~\citep{spearman1904proof} for rank correlations between CEFR ordinal ranks and median CT scores per cluster.

To contextualize FCM performance, two widely-used clustering algorithms (MiniBatchKMeans and DBSCAN) were evaluated under identical conditions on a stratified training subset.


\section{Results}
\label{sec:results}

\subsection{Model Performance and Generalization}

\begin{figure*}[htbp]
	\centering
	\includegraphics[width=\linewidth]{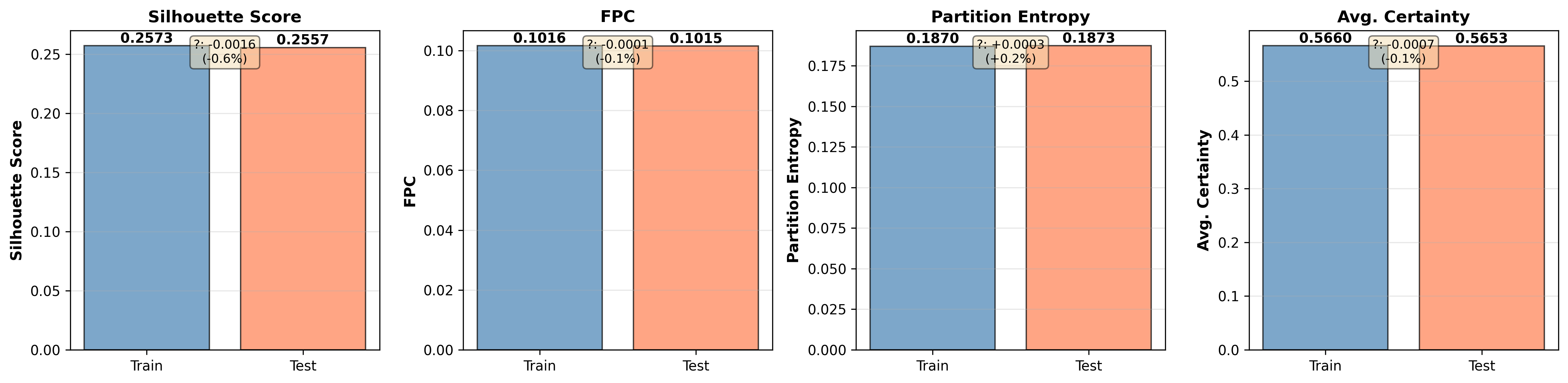}
	\caption{Train-test performance comparison across four key metrics. Bars represent metric values for training (blue) and test (coral) sets. Difference labels show absolute and percentage changes, demonstrating minimal generalization gap.}
	\label{fig:train_test_comparison}
\end{figure*}

The Fuzzy C-Means algorithm converged on the training set ($n = 1{,}606{,}596$) with Silhouette Score~\citep{rousseeuw1987silhouettes} = 0.2573, FPC = 0.6099, Partition Entropy = 1.1219, and Average Certainty = 0.5660. The test set ($n = 401{,}650$) achieved nearly identical metrics: Silhouette = 0.2557 ($\Delta = -0.0016$, $-0.6\%$), FPC = 0.6091 ($\Delta = -0.0008$, $-0.1\%$), Partition Entropy = 1.1238 ($\Delta = +0.0019$, $+0.2\%$), and Average Certainty = 0.5653 ($\Delta = -0.0007$, $-0.1\%$), see Figure~\ref{fig:train_test_comparison}.

Five-fold cross-validation yielded Silhouette: mean = 0.2631, $\sigma^2 = 0.000080$; FPC: mean = 0.6139, $\sigma^2 = 0.000040$.

\subsection{Centroid Profiles and Statistical Validation}

Table~\ref{tab:centroids} presents the ordered cluster centroids via $S_j$ criterion. Profiles show progression from A1 ($S_j = 2.70$, near-zero proficiency except User Interactivity 1.01) to C2 ($S_j = 27.98$, near-maximal scores across dimensions).

\begin{table}[htbp]
	\centering
	\caption{Ordered cluster centroids with CEFR level assignments (values represent mean CT scores on 0--4 scale)}
	\label{tab:centroids}
	\scriptsize
	\setlength{\tabcolsep}{3pt}
	\begin{tabular}{@{}l@{\hspace{2.5pt}}ccccccccccc@{}}
		\toprule
		\textbf{CEFR} & \textbf{Abs.} & \textbf{Par.} & \textbf{Log.} & \textbf{Syn.} &
		\textbf{Flow} & \textbf{User} & \textbf{Data} & \textbf{Math} & \textbf{Mot.} & \textbf{$S_j$} \\
		\midrule
		A1 & 0.12 & 0.18 & 0.09 & 0.15 & 0.52 & 1.01 & 0.24 & 0.08 & 0.31 & 2.70 \\
		A2 & 0.89 & 1.02 & 0.65 & 0.88 & 1.76 & 1.45 & 1.34 & 0.43 & 1.15 & 9.57 \\
		B1 & 1.58 & 1.89 & 1.35 & 1.58 & 2.45 & 1.68 & 2.18 & 0.82 & 1.92 & 15.45 \\
		B2 & 2.21 & 2.64 & 2.01 & 2.18 & 2.85 & 1.84 & 2.89 & 1.21 & 2.55 & 20.38 \\
		C1 & 2.78 & 3.18 & 2.62 & 2.71 & 3.15 & 1.95 & 3.45 & 1.59 & 3.04 & 24.47 \\
		C2 & 3.29 & 3.61 & 3.19 & 3.19 & 3.38 & 2.03 & 3.87 & 1.96 & 3.46 & 27.98 \\
		\bottomrule
		\addlinespace[0.5ex]
		\multicolumn{11}{@{}l@{}}{%
			\tiny
			\parbox{\columnwidth}{%
				\textit{Note:} Abs. = Abstraction, Par. = Parallelization,
				Log. = Logic, Syn. = Synchronization, Flow = Flow Control,
				User = User Interactivity, Data = Data Representation,
				Math = Math Operators, Mot. = Motion Operators.}%
		}
	\end{tabular}
\end{table}

\begin{figure}[htbp]
	\centering
	\includegraphics[width=0.85\linewidth]{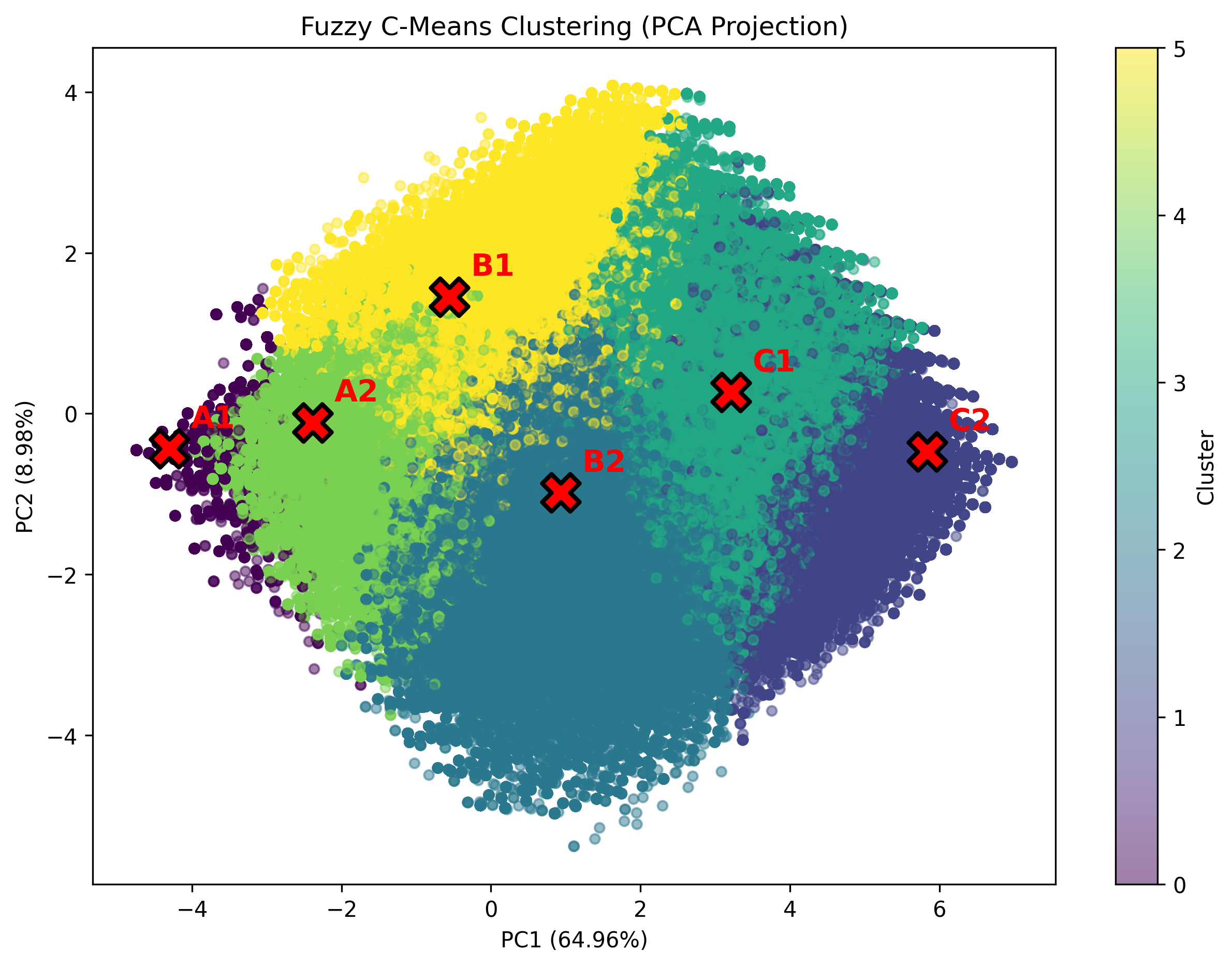}
	\caption{PCA projection of training set observations colored by FCM cluster assignment. Red X markers indicate CEFR-ordered centroids. Substantial overlap between adjacent levels reflects natural skill progression continuum.}
	\label{fig:pca_clusters}
\end{figure}

Principal Component Analysis (PCA)~\citep{jolliffe2016principal} visualization, see Figure~\ref{fig:pca_clusters}, explains 73.89\% variance (PC1: 64.90\%, PC2: 8.99\%), confirming ordinal arrangement with substantial overlap between adjacent levels.

Kruskal-Wallis test~\citep{mckight2010kruskal}: $H = 7{,}609{,}964.74$ ($p < 0.001$). All pairwise Mann-Whitney U tests~\citep{mcknight2010mann}: $p < 0.001$. Global correlations: Kendall's~\citep{kendall1938new} $\tau = 1.000$, Spearman's~\citep{spearman1904proof} $\rho = 1.000$ ($p < 0.001$). Dimension-specific correlations, as described in Table~\ref{tab:dimension_correlations}, show Synchronization ($\tau = 0.966$), Flow Control ($\tau = 0.894$), and Logic ($\tau = 0.873$) as the strongest ordinal indicators.

\begin{table}[ht]
	\centering
	\caption{Dimension-specific rank correlations with CEFR level}
	\label{tab:dimension_correlations}
	\scriptsize
	\begin{tabularx}{\linewidth}{lXXXX}
		\toprule
		\textbf{CT Dimension} & \textbf{$\tau$} & \textbf{$p_\tau$} & \textbf{$\rho$} & \textbf{$p_\rho$} \\
		\midrule
		Synchronization & 0.966 & $<0.001$ & 1.000 & $<0.001$ \\
		Flow Control & 0.894 & $<0.001$ & 0.986 & $<0.001$ \\
		Logic & 0.873 & $<0.001$ & 0.971 & $<0.001$ \\
		Abstraction & 0.845 & $<0.001$ & 0.957 & $<0.001$ \\
		Data Representation & 0.822 & $<0.001$ & 0.943 & $<0.001$ \\
		Parallelization & 0.810 & $<0.001$ & 0.929 & $<0.001$ \\
		Motion Operators & 0.789 & $<0.001$ & 0.914 & $<0.001$ \\
		Math Operators & 0.756 & $<0.001$ & 0.886 & $<0.001$ \\
		User Interactivity & 0.623 & $<0.001$ & 0.771 & $<0.001$ \\
		\bottomrule
	\end{tabularx}
\end{table}

\paragraph{Classification Stability}

Sensitivity analysis confirms the robustness of the enhanced classification system. Varying $\tau_{clear}$ by $\pm0.05$ shifts Clear classifications by 6--7 percentage points, while $\tau_{trans}$ variations produce proportionate changes in Transition detection (approximately 2 percentage points per 0.025 increment). Importantly, at the selected $\tau_{trans}=0.15$, over 75\% of detected transitions involve adjacent CEFR levels, supporting the interpretation that transitional classifications reflect genuine developmental states rather than arbitrary boundary effects.

\subsection{Enhanced Classification System Results}

\begin{figure*}[htbp]
	\centering
	\includegraphics[width=0.9\linewidth]{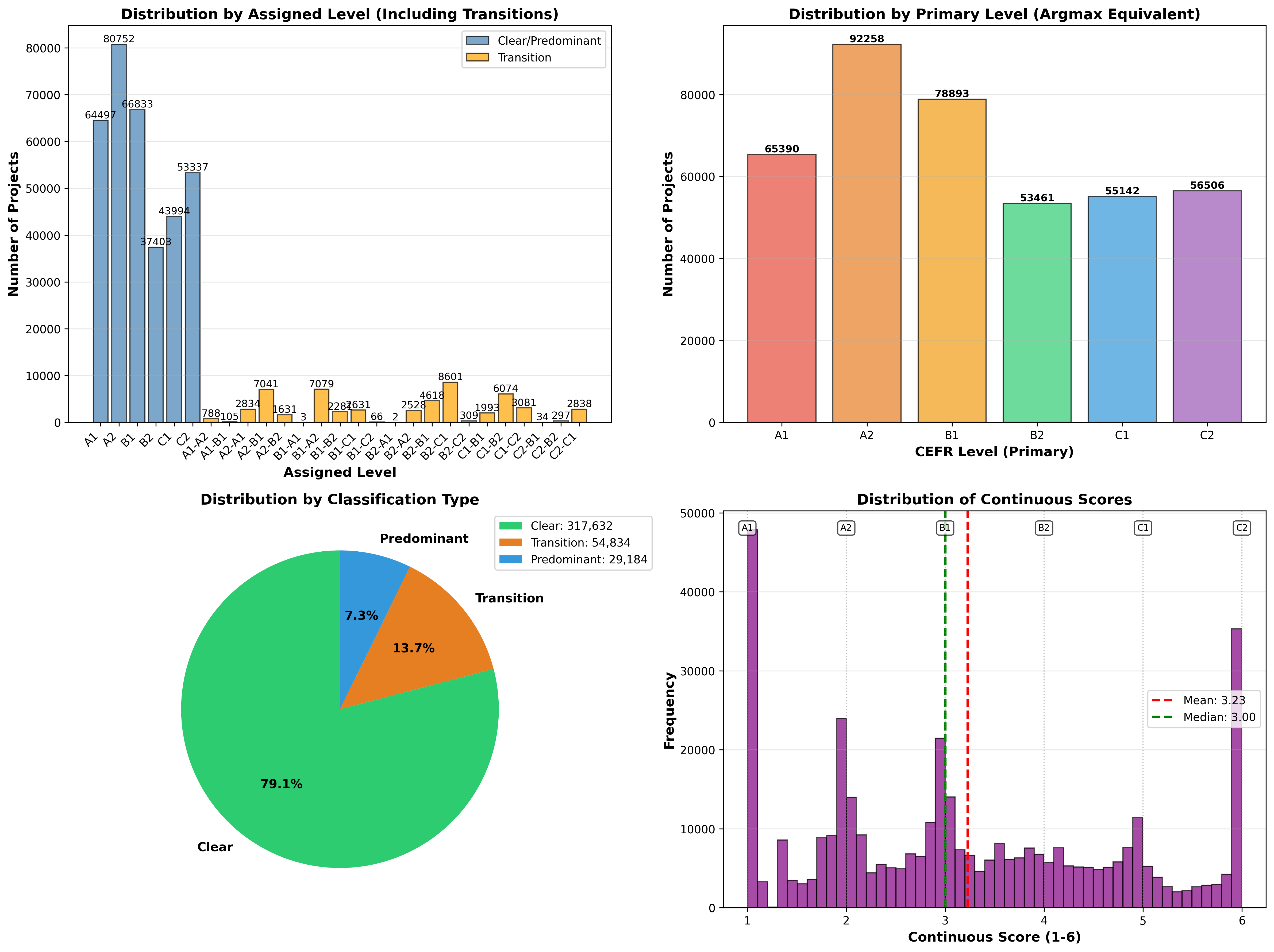}
	\caption{Enhanced classification system results on test set. (Top left) Distribution by assigned level including transition states (e.g., A1-A2, B2-C1). (Top right) Distribution by primary level (\texttt{argmax} equivalent). (Bottom left) Pie chart showing classification type proportions. (Bottom right) Histogram of continuous scores (1--6 scale) with CEFR level markers.}
	\label{fig:classification_distribution}
\end{figure*}

Test set classification distribution: 79.1\% Clear, 13.7\% Transition, 7.3\% Predominant, see Figure~\ref{fig:classification_distribution}. Primary level distribution: A1--A2 (39.3\%), B1 (19.6\%), B2 (13.3\%), C1--C2 (27.8\%). Notable transition populations: A2-B1 (7,041, 1.8\%), B2-C1 (8,601, 2.1\%), C1-C2 (3,081, 0.8\%).

These transition populations represent students in what Vygotsky~\citep{Vygotsky1978Mind} termed the ``zone of proximal development''---actively developing skills between adjacent levels, where technology-supported scaffolding is most effective. Where technology-supported scaffolding is most effective. The substantial B2-C1 transition group (8,601 learners, 2.1\%) exemplifies the cognitive load challenge: these students must simultaneously integrate Logic, Synchronization, and Data Representation, a qualitative shift from earlier levels where skills develop more independently. This is precisely where automated flagging enables targeted human intervention.

The continuous score distribution (mean = 3.23, median = 3.00, SD = 1.55) exhibits multiple modes corresponding to the six CEFR levels, with substantial density between integer values.

\begin{figure*}[htbp]
	\centering
	\includegraphics[width=\linewidth]{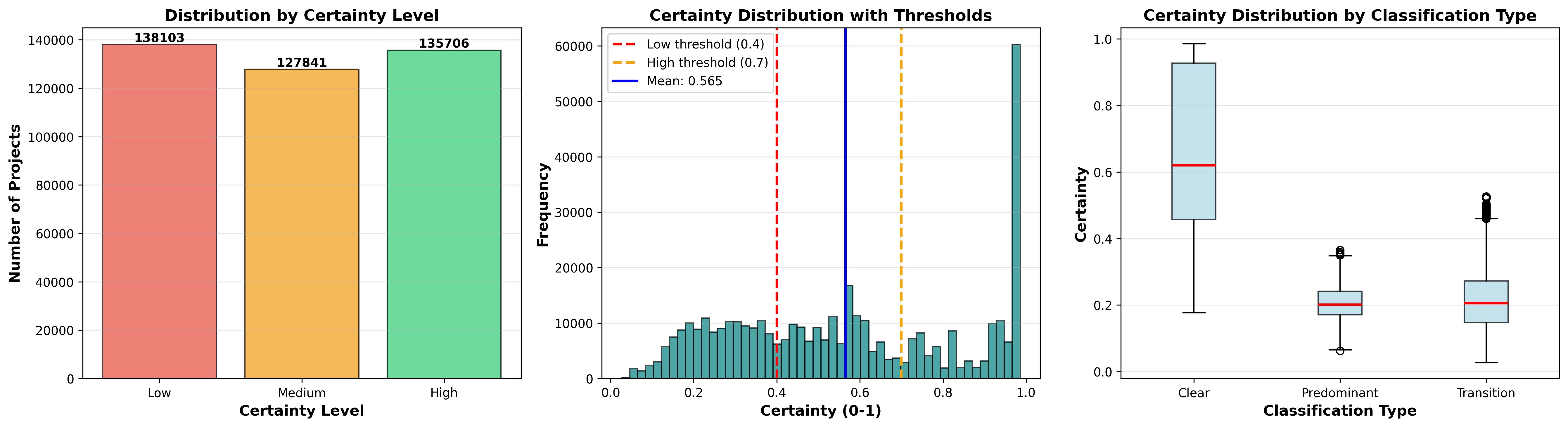}
	\caption{Certainty analysis on test set. (Left) Distribution by categorical certainty level (Low/Medium/High). (Center) Histogram of continuous certainty values with threshold markers and mean indicator. (Right) Box plots showing certainty distributions stratified by classification type.}
	\label{fig:certainty_analysis}
\end{figure*}

Certainty analysis reveals a relatively balanced distribution across levels: Low (34.4\%), Medium (31.8\%), High (33.8\%)---Figure~\ref{fig:certainty_analysis}.

\begin{figure*}[htbp]
	\centering
	\includegraphics[width=0.9\linewidth]{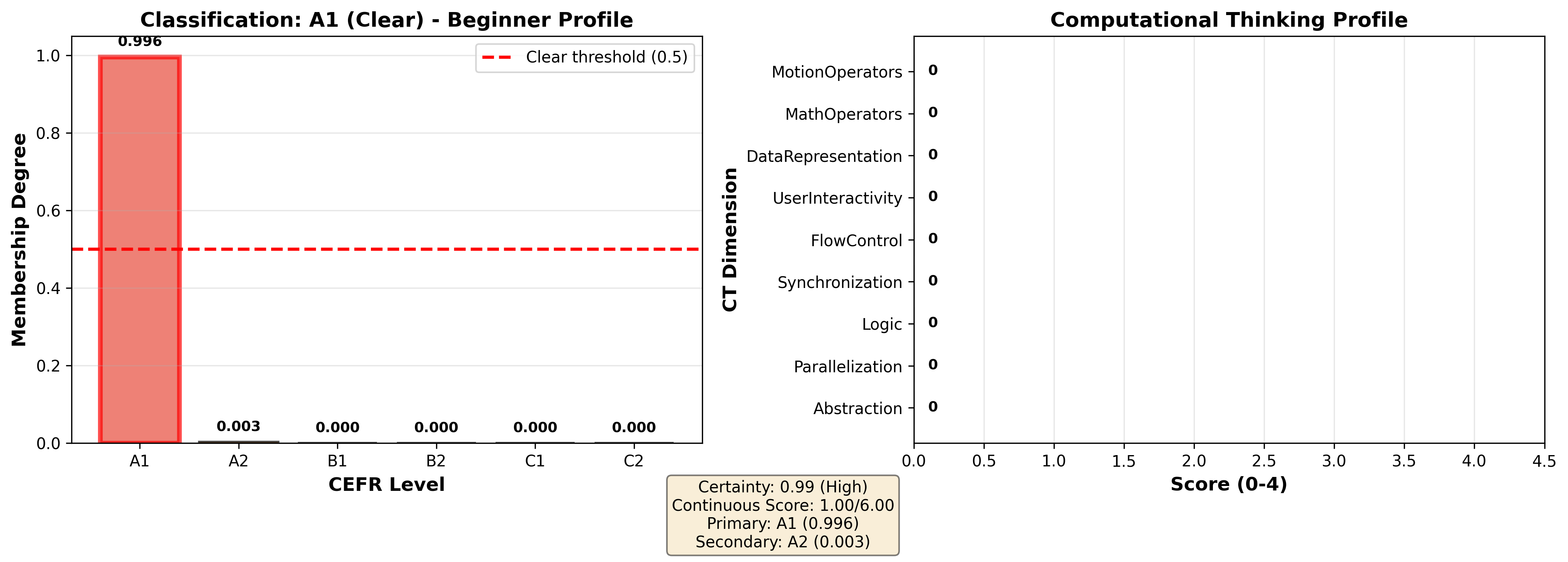}
	\caption{Example A1 (Clear) classification. Complete beginner profile with all CT dimensions at zero. Primary membership = 0.996, certainty = 0.99, continuous score = 1.00/6.00.}
	\label{fig:example_beginner}
\end{figure*}

\begin{figure*}[htbp]
	\centering
	\includegraphics[width=0.9\linewidth]{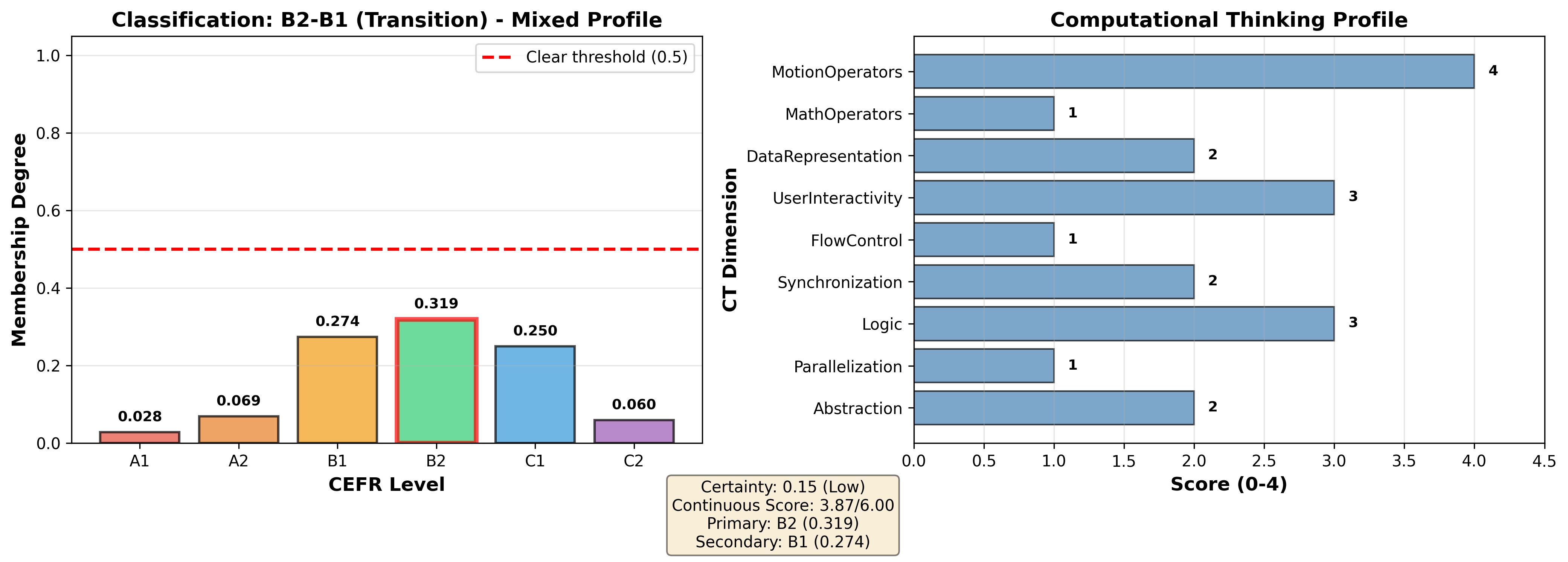}
	\caption{Example B2-B1 (Transition) classification. Mixed skill profile with heterogeneous CT scores. Membership balanced between B2 (0.319) and B1 (0.274), certainty = 0.15, continuous score = 3.87/6.00.}
	\label{fig:example_transition}
\end{figure*}

\begin{figure*}[htbp]
	\centering
	\includegraphics[width=0.9\linewidth]{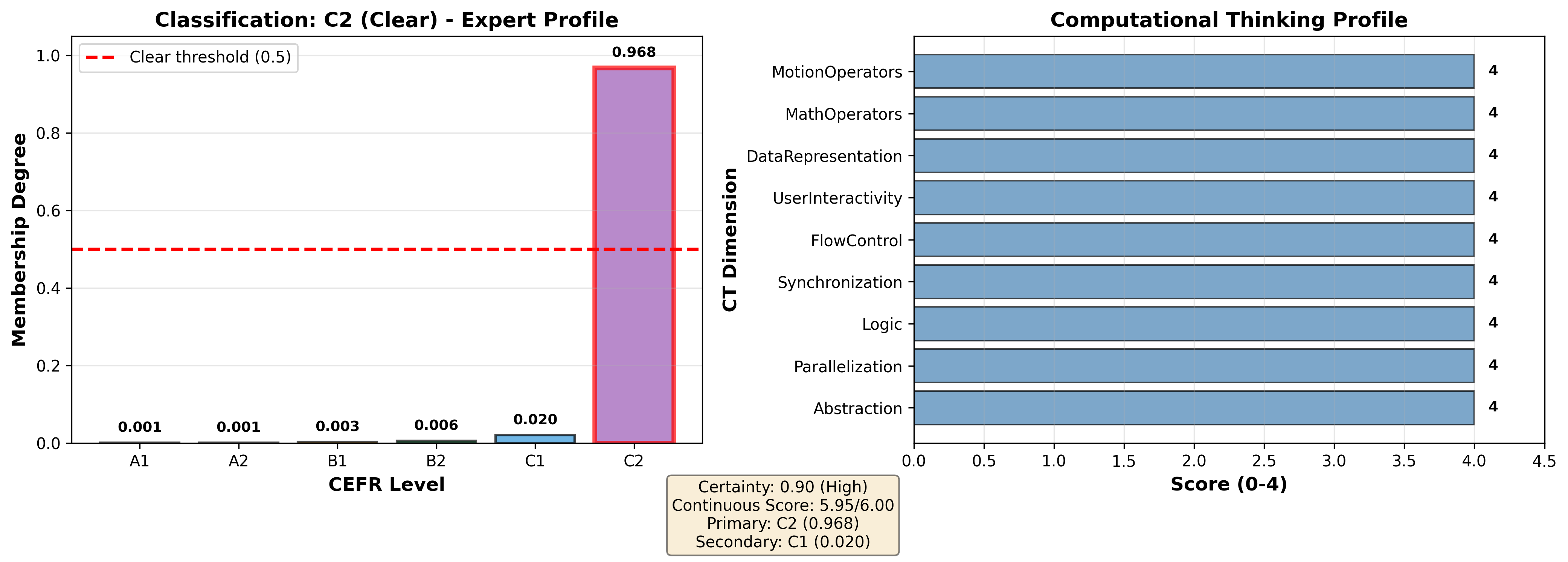}
	\caption{Example C2 (Clear) classification. Expert profile with maximal scores (4) across all nine CT dimensions. Primary membership = 0.968, certainty = 0.90, continuous score = 5.95/6.00.}
	\label{fig:example_expert}
\end{figure*}

Representative examples in Figures~\ref{fig:example_beginner}, \ref{fig:example_transition}, and \ref{fig:example_expert}: A1 beginner (all zeros, certainty=0.99), B2-B1 transition (mixed skills, certainty=0.15), C2 expert (all fours, certainty=0.90).

\subsection{Baseline Algorithm Comparison}

MiniBatchKMeans~\citep{macqueen1967} ($k=6$) achieved Silhouette Score~\citep{rousseeuw1987silhouettes} = 0.293 (vs. 0.260 for FCM) but dramatically lower Average Proportion of Non-neighbors (APN)~\citep{ben2001stability} (APN = 0.023 vs. 0.097), indicating excessively rigid cluster boundaries that fail to capture transitional states. DBSCAN~\citep{schubert_dbscan_2017} produced fragmented output (62 clusters, 11.3\% noise) with negative Silhouette Score ($-0.156$), rendering it unsuitable for ordinal skill classification.

These results underscore FCM's unique capacity to balance cluster separation with membership overlap, essential for modeling educational progression. Unlike MiniBatchKMeans, which maximizes cluster separation at the expense of membership overlap, FCM preserves soft boundaries essential for representing transitional competencies.


\section{Discussion}

\label{sec:discussion}

This study demonstrates that Fuzzy C-Means clustering with ordinal $S_j$ mapping provides a robust, interpretable framework for CEFR-aligned programming skill assessment at scale. We discuss the key implications of our findings for educational practice, methodological considerations, and the broader context of computational thinking assessment.

\paragraph{The B2 Bottleneck: A Target for Curriculum Intervention}

The bimodal distribution reveals concentration at beginner (A1--A2: 39.3\%) and advanced (C1--C2: 27.8\%) stages with a pronounced deficit at B2 (13.3\%). This ``B2 bottleneck'' represents a critical pedagogical finding: it identifies precisely where learners struggle most in their progression from novice to proficient programmer.

Analysis of centroid profiles suggests why B2 presents such difficulty: learners at this level must simultaneously integrate Logic (centroid value 2.01), Synchronization (2.18), and Data Representation (2.89)---a cognitive load substantially greater than earlier levels where skills develop more independently. This is precisely where automated flagging enables targeted human intervention, see Table~\ref{tab:feedback_strategies}.

\begin{table}[htbp]
	\centering
	\scriptsize
	\caption{Automated feedback strategies by classification}
	\label{tab:feedback_strategies}
	\begin{tabularx}{\linewidth}{llX}
		\toprule
		\textbf{Level} & \textbf{Certainty} & \textbf{Suggested Intervention} \\
		\midrule
		A1--A2 & High & ``How could you make this character repeat its dance?'' (Flow Control) \\
		B1 & High & ``What if you needed to remember the player's score?'' (Data) \\
		B2 & Any & Trigger synchronization challenge activity \\
		B2--C1 & Low & Flag for instructor review + peer mentoring \\
		C1--C2 & High & Can you spot repeated patterns to simplify? \\
		\bottomrule
	\end{tabularx}
\end{table}

The dimension-specific correlations further inform curriculum sequencing. Synchronization ($\tau=0.966$), Flow Control ($\tau=0.894$), and Logic ($\tau=0.873$) emerge as the strongest ordinal indicators, suggesting these competencies should anchor advanced curricula and serve as gatekeeping assessments for C-level certification. Conversely, User Interactivity's early plateauing (1.01--2.03 across all levels) indicates this dimension differentiates primarily among beginners and should be front-loaded in introductory instruction rather than emphasized throughout.

\paragraph{Certainty as a Trigger for Human Intervention}

The relatively balanced certainty distribution---Low (34.4\%), Medium (31.8\%), High (33.8\%)---has important implications for deploying this framework in educational practice. Approximately one-third of classifications achieve high certainty ($\geq 0.7$), indicating these assessments can reliably support automated formative feedback without instructor oversight. The continuous score and CEFR level assignment for these learners can be communicated directly, with dimension-specific recommendations generated automatically.

Conversely, the 34.4\% of classifications with low certainty ($< 0.4$) warrant human review. These cases often involve learners with heterogeneous skill profiles---strong in some dimensions, weak in others---that defy simple categorization. For these students, the membership vector itself becomes pedagogically valuable: an instructor reviewing a learner with memberships split between B1 (0.35), B2 (0.30), and A2 (0.25) gains insight into the specific developmental pattern that a single level assignment would obscure. Consider a concrete scenario: a student flagged as ``B2-B1 Transition'' shows strong Motion Operators (4) but weak Synchronization (2). The teacher, alerted via the flagging queue, recognizes this as a student who creates visually impressive animations but struggles to coordinate multiple sprites---a common pattern. Rather than generic ``B2'' feedback, the instructor designs a targeted activity combining the student's motion strengths with explicit synchronization challenges.

This certainty-driven triage enables efficient allocation of instructor attention: by automating feedback for the 33.8\% of High-certainty cases, instructors can dedicate their full attention to Low-certainty learners where qualitative interpretation and personalized guidance are essential---effectively multiplying the impact of limited human resources.

The threshold selection further supports this triage approach. Sensitivity analysis reveals that $\tau_{low}=0.4$ and $\tau_{high}=0.7$ align with distributional terciles, ensuring approximately equal-sized groups for each certainty level. This prevents the ``Low certainty'' category from becoming either too restrictive (missing genuinely ambiguous cases) or too inclusive (overwhelming instructors with borderline classifications). Educational institutions can adjust these thresholds based on available instructor capacity: settings with limited human review resources might raise $\tau_{low}$ to focus attention on only the most ambiguous cases.

\paragraph{Operationalization: A Conceptual Dashboard}

The framework's outputs translate directly into actionable interface elements: (1) a \textit{flagging queue} highlighting Low-certainty students for immediate review, (2) a \textit{progress timeline} showing $S_{\text{cont}}$ evolution across submissions---visualizing growth even within a single CEFR band, and (3) \textit{transition alerts} notifying instructors when students enter ambiguous developmental states. This design prioritizes instructor agency: the algorithm surfaces patterns, but pedagogical judgment remains human.

\paragraph{Implications for Professional Contexts}

This framework, although developed for educational assessment, could also be applied to professional training and hiring. A CEFR-aligned programming classification system would enable transparent hiring by allowing candidates to present verified CEFR programming credentials similar to language proficiency certificates, guide training programs by aligning corporate learning content with employees' assessed CEFR levels, and help organizations benchmark capabilities by aggregating CEFR distributions to identify strengths and skill gaps. However, applying this framework in professional settings requires validating that CEFR levels inferred from Scratch projects predict performance on real-world programming tasks, which involve collaboration, debugging, and domain-specific knowledge not yet captured by this approach.

\paragraph{Limitations}

Key limitations include: (1) \textbf{Scratch-specific applicability}---adaptation to text-based languages requires domain-specific feature engineering; (2) \textbf{Static assessment}---longitudinal tracking would reveal developmental trajectories; (3) \textbf{Cultural context}---the dataset is predominantly English-speaking, requiring cross-cultural validation; (4) \textbf{Construct scope}---Dr.Scratch metrics omit code quality dimensions (elegance, maintainability, creativity). However, this limitation also presents an opportunity: by automating assessment of functional CT dimensions, the framework frees instructor time to focus on evaluating the creative and aesthetic qualities that currently require human judgment.


\section{Future Work}
\label{sec:future}

Building on the foundation established in this study, we identify several promising directions for extending CEFR-aligned programming skill assessment into institutional practice and research.

\paragraph{Formal Credentialing and Institutional Integration}

The alignment with CEFR---an internationally recognized framework already embedded in educational policy across Europe and beyond---positions this work for direct integration into formal credentialing systems. Priority development directions include:

\begin{itemize}[noitemsep,topsep=0pt]
	\item \textbf{Standardized certification exams:} Develop psychometrically validated assessments that certify demonstrated CEFR programming levels, analogous to language proficiency certificates.
	\item \textbf{Blockchain-backed digital badges:} Issue tamper-proof credentials for achieved levels, enabling portable verification across educational institutions and employers.
	\item \textbf{Competency-based progression:} Shift from seat-time to mastery-based advancement, using $S_{\text{cont}}$ thresholds as promotion criteria. The B2 bottleneck data could justify institutional funding for specialized ``bridge courses'' targeting the Logic-Synchronization-Data integration challenge.
	\item \textbf{DigCompEdu alignment:} Integrate programming CEFR levels into the existing Digital Competence of Educators framework as an automatic certification module, enabling teachers to validate their digital competencies through their own Scratch projects and demonstrate competence to assess student work at each CEFR level.
\end{itemize}

For professional contexts, CEFR-aligned programming credentials could transform hiring and training practices: candidates present verified proficiency levels comparable to language certifications, organizations benchmark workforce capabilities through aggregated CEFR distributions, and training programs target specific level transitions identified through assessment.

\paragraph{Research Extensions}
Priority research directions include: longitudinal tracking to investigate transition duration and progression patterns; extension to text-based languages (Python, JavaScript) through language-agnostic CT metrics and AST-based analysis; multimodal assessment integrating portfolio analysis, debugging behavior, and collaborative coding; and cross-cultural validation across Scratch's 150+ country user base.

\paragraph{Educational Technology Integration}
Develop adaptive learning systems leveraging real-time CEFR classification for dynamic scaffolding, personalized learning paths, and automated formative feedback. Release open-source infrastructure including a Python/R package and public web platform for instant project classification.


\section{Conclusions}
\label{sec:conclusions}

This work presents a CEFR-aligned framework for automated programming skill assessment using Fuzzy C-Means clustering on \Ntotal{} Scratch projects. Our approach addresses three gaps in the current literature.

First, we bridge discrete assessment and continuous development with pedagogical utility. Unlike existing tools that force learners into categorical bands, our fuzzy membership vectors capture proximity to multiple proficiency levels simultaneously. The enhanced classification system identifies 13.7\% of learners as actively transitioning between adjacent levels---information essential for targeted pedagogical intervention that would be lost in traditional hard clustering approaches. The certainty metric further operationalizes when automated feedback suffices versus when instructor review is warranted.

Second, we establish standardized competency measurement. By mapping clusters to the internationally recognized CEFR framework via the $S_j$ ordering criterion, we enable direct comparability across educational contexts. This data-driven ordinal mapping is applicable beyond Scratch to any domain requiring alignment between unsupervised clustering and established proficiency scales.

Third, we reconcile statistical and pedagogical validity. Although internal validity indices favor $k=2$ clusters, we demonstrate that domain-aligned six-cluster solutions remain statistically defensible when all pairwise comparisons are significant (Mann--Whitney $p < 0.001$) and rank correlations confirm monotonic ordering ($\tau = 1.0$).

The empirical validation establishes benchmark metrics (Silhouette $\approx 0.257$, Average Certainty $\approx 0.566$) with robust generalization ($\Delta$ Silhouette $= -0.6\%$ train-to-test). Key findings include the identification of Synchronization, Flow Control, and Logic as strongest progression indicators ($\tau > 0.87$), and a B2-level bottleneck (13.3\%) suggesting targets for pedagogical intervention.

By demonstrating that fuzzy clustering can impose educationally meaningful ordinal structure while preserving soft boundaries essential for capturing transitional learning states, this work establishes a foundation for standardized, continuous, and uncertainty-aware competency assessment in programming education.


\bibliographystyle{apalike}
\bibliography{CSEDU2026}

\end{document}